\documentclass[dvipdfmx]{pasj01}
\draft

\begin{document} 
\Received{}
\Accepted{}

\title{The Galactic Center Lobe Filled with Thermal Plasma}

\author{
  Halca \textsc{Nagoshi}\altaffilmark{1},
  Yuzo \textsc{Kubose}\altaffilmark{1},
  Kenta \textsc{Fujisawa}\altaffilmark{2*},
  Kazuo \textsc{Sorai}\altaffilmark{3, 4, 5, 6},
  Yoshinori \textsc{Yonekura}\altaffilmark{7},
  Koichiro \textsc{Sugiyama}\altaffilmark{8},
  Kotaro \textsc{Niinuma}\altaffilmark{1},
  Kazuhito \textsc{Motogi}\altaffilmark{1},
  and
  Takahiro \textsc{Aoki}\altaffilmark{2}}

\altaffiltext{1}{Graduate School of Sciences and Technology for Innovation, Yamaguchi University, Yoshida 1677-1, Yamaguchi-city, Yamaguchi 753-8512, Japan}
\altaffiltext{2}{The Research Institute for Time Studies, Yamaguchi University, Yoshida 1677-1, Yamaguchi-city, Yamaguchi 753-8511, Japan}
\altaffiltext{3}{Department of Physics, Faculty of Science, Hokkaido University, Kita 10 Nishi 8, Kita-ku, Sapporo 060-0810, Japan}
\altaffiltext{4}{Department of Cosmosciences, Graduate School of Science, Hokkaido University, Kita 10 Nishi 8, Kita-ku, Sapporo 060-0810, Japan}
\altaffiltext{5}{Division of Physics, Faculty of Pure and Applied Sciences, University of Tsukuba, 1-1-1 Tennodai, Tsukuba, Ibaraki 305-8571, Japan}
\altaffiltext{6}{Tomonaga Center for the History of the Universe, University of Tsukuba, 1-1-1 Tennodai, Tsukuba, Ibaraki 305-8571, Japan}
\altaffiltext{7}{Center for Astronomy, Ibaraki University, 2-1-1 Bunkyo, Mito, Ibaraki 310-8512, Japan}
\altaffiltext{8}{Mizusawa VLBI Observatory, National Astronomical Observatory of Japan, 2-21-1 Osawa, Mitaka, Tokyo 181-8588, Japan}
\email{kenta@yamaguchi-u.ac.jp}

\KeyWords{Galaxy: center --- radio lines: ISM --- HII regions}

\maketitle

\begin{abstract}
An observational result of a radio continuum and H92$\alpha$ radio recombination line of the Galactic Center Lobe (GCL), using the Yamaguchi 32 m radio telescope, is reported. The obtained spatial intensity distribution of the radio recombination line shows two distinctive ridge-like structures extending from the galactic plane vertically to the north at the eastern and western sides of the galactic center, which are connected to each other at a latitude of $1.2^{\circ}$ to form a loop-like structure as a whole. This suggests that most of the radio continuum emission of the GCL is free-free emission, and that the GCL is filled with thermal plasma. The east ridge of the GCL observed with the radio recombination line separates 30 pc from the radio arc, which has been considered as a part of the GCL, but coincides with a ridge of the radio continuum at a galactic longitude of $0^{\circ}$. The radial velocity of the radio recombination line is found to be between $-4$ and $+10$ km s$^{-1}$ across the GCL. This velocity is much smaller than the one expected from the galactic rotation, and hence indicates that the GCL exists apart from the galactic center. These characteristics of the GCL suggest that the long-standing hypothesis that the GCL was created by an explosive activity in the galactic center is unlikely, but favor that the GCL is a giant HII region.
\end{abstract}

\section{Introduction}
The Galactic Center Lobe (GCL) is a radio structure extending for one degree, located near the galactic center. It was discovered with 10 GHz continuum observations with the 45 m radio telescope at Nobeyama (Sofue \& Handa 1984, Sofue 1985). The radio structure of the GCL has been studied with the Effelsberg 100 m telescope (Reich et al. 1990 a, b), Parkes 64 m telescope (Haynes et al. 1992), GreenBank 100 m telescope, and the VLA (Law et al. 2008 a, b). The GCL shows a distinctive structure in the image, after which it is named; it has two ridge-like structures of radio continuum emission located at the eastern and western sides of the galactic center that extend from the galactic plane toward the galactic-north and they appear to be connected at a galactic latitude of roughly plus one degree, forming a dome-like shape overall. It is argued by Sofue (1985) and in following studies that the east and west ridges are connected to the radio arc near the galactic plane and to the star forming region Sgr C, respectively, both of which are known to be physically situated in the galactic center region (Lasenby et al. 1989). The emission of the radio arc is synchrotron radiation from high energy electrons in a strong magnetic field of up to 1 mG (Yusef-Zadeh et al. 1984). On the basis of it, Reich et al. (1987) argued that the radiation of the entire GCL would be synchrotron radiation. Law et al. (2009), however, detected radio recombination lines of hydrogen, which are emitted from thermal plasma, from the GCL. Their detection suggests that the radio continuum of the GCL is, at least partially, free - free emission of thermal plasma. H$\alpha$ emission was also observed at near the top of the GCL (Gaustad et al. 2001). An infrared counterpart structure of the GCL has been detected with MSX and Spitzer (Bland-Hawthorn \& Cohen 2003, Arendt et al. 2008, Yusef-Zadeh et al. 2009). The fact that the observed infrared radiation is in thermal dust origin supports the thermal nature of the GCL.

The GCL extends vertically from the galactic plane for 150 pc if it is in the galactic center. Given that the radio continuum is bright at the east and west edges, the spatial density distribution of the GCL must be shell-like, i.e., low in the center and high at the edge (Sofue 1985). On the basis of this structure, formation models of the GCL have been proposed in which association with the activity of the galactic center region is argued (e.g., Crocker et al. 2011). Bland-Hawthorn \& Cohen (2003) discussed similarity of the GCL to outflows in starburst galaxies such as NGC 3079. Law et al. (2009) argued that the GCL is an ionized region by photons from large star clusters. Sofue (1985) argued that the poloidal magnetic field that penetrates the galactic plane is twisted by the galactic rotation, and that the GCL is a structure ejected from the galactic plane by the energy accumulated in the magnetic field. The radiation mechanism, structure, and motion of the GCL are keys to validate the models of its formation mechanisms.

Most of the previous observations of the GCL have been made with radio continuum. Radio continuum observations of objects close to the galactic center are affected by many objects in the line of sight that are unrelated to the target object, and that makes study of the target difficult. If the GCL contains thermal plasma, it would emit radio recombination lines of hydrogen as well as radio continuum by free-free emission. The spatial intensity distributions of the radio continuum and the recombination line should show the same structures if the GCL is filled with thermal plasma. In other words, observations of radio recombination lines enable us to selectively observe thermal plasma, and to filter out synchrotron radiation emitted by unrelated objects in the line of sight. Furthermore, the velocity of the thermal plasma can be investigated with radio recombination-line observations. Lockman \& Gordon (1973, H159$\alpha$, 1.6 GHz) and Pauls \& Mezger (1975, H109$\alpha$, 5 GHz) observed radio recombination lines at positions near the GCL and showed that the radial velocity was close to 0 km s$^{-1}$. Law et al. (2009) observed the entire region of the GCL at H109$\alpha$ (5 GHz) using the HCRO 25 m, and a part of the GCL at H106-113$\alpha$ (5 GHz) using the GBT 100 m. Alves et al. (2015) conducted radio recombination line observations for a wide area of the galactic plane, including the GCL at H166-168$\alpha$ (1.4 GHz), using the Parkes 64 m (HI Parkes All-Sky Survey, HIPASS). These observations show that the spatial intensity distribution of radio recombination lines, that is, the distribution of thermal plasma, is loop-like with two ridges in the east and west, being similar to the shape of the GCL observed in radio continuum, and that the radial velocity is close to 0 km s$^{-1}$. In these past observations, however, the beam size was large and the angular resolution was insufficient (HCRO, Parkes) or the observation area was limited (GBT) to investigate the relation between the radio recombination line and the continuum of the GCL in detail. It is required to observe the whole area of the GCL with high angular resolution in radio continuum and radio recombination line.

We made observations of the whole area of the GCL with radio continuum at 8.4 GHz and H92$\alpha$ radio recombination line with the Yamaguchi 32m radio telescope in order to investigate the radiation mechanism, structure, and the formation mechanism of the GCL. In this paper, we describe the observations in section 2 and summarize the results in section 3. In section 4, we discuss the structure, radiation mechanism, physical parameters, and the formation mechanism of the GCL based on the observation results. Our results and discussion are summarized in section 5.

\section{Observations}
\subsection{Radio continuum observations}
The continuum observations were made with the Yamaguchi 32 m radio telescope (Fujisawa et al. 2002). Continuum emission at center frequency of 8380 MHz with a bandwidth of 400 MHz in LHCP and RHCP were simultaneously observed. A cooled low-noise receiver with a system-noise temperature of 45 - 50 K was used. The FWHM of the beam was $4.3'$. The main beam efficiency was $0.70$. The sidelobe level was $-13$ dB at $6.5'$ with respect to the beam center, and was lower than $-30$ dB outside $30'$ from the beam center. The squint of the beams of two polarization is smaller than $0.1'$.

The observed area was $-2^{\circ} 33' \le l \le +2^{\circ} 27'$ and $-2^{\circ} 25' \le b \le +2^{\circ} 35'$; the center of the observed area is slightly off the origin of the galactic coordinates. The sky was scanned at a rate of $5^{\circ}$ per minute in the galactic latitudinal direction. The scan was repeated $301$ times with 1 arc minute shift in longitude in between. The power was measured continuously during the scan at a rate of 1 sample per $\sim 0.01$ seconds. The data were time-averaged over $0.2$ seconds to obtain one piece of data. This resulted in a total of $301 \times 301$ data points on 1 arc minute grid in latitude and longitude covering a $5 \times 5^{\circ}$ area. The pointing error was 1 arc minute at maximum. One observation took about 5.5 hours for one map. Observations were done twice on 2006 April 16 and 17 for the same area. The weather was clear on the two observation days. The integration time for each point was effectively 0.8 seconds after summation of dual polarization and two days observation.

The conversion from the measured power to antenna temperature was performed as follows. System noise temperatures were measured using a calibrated noise source before and after the observation to the blank sky at elevation angles from $7^{\circ}$ to $90^{\circ}$, and the system noise temperature was modeled for the blank sky as a function of the elevation angle. The modeled system noise temperature at each elevation angle was subtracted from the observed value, and the excess antenna temperature due to radio sources was obtained. This antenna temperature in each scan had an offset caused by a slight variation of the system gain and atmospheric absorption. In order to eliminate this gain variation, 5 points at both ends of one scan (data of latitude $-2^{\circ} 25'$ to $-2^{\circ} 21'$, and $+2^{\circ} 31'$ to $+2^{\circ} 35'$) were used to obtain zero-level baseline, which was then subtracted from the observed data. Given that there are no bright radio sources near latitudes of $\pm 2.5^{\circ}$ in the image of the Effelsberg 11 cm observation (Reich et al. 1990a, b), the radio brightness distribution at these latitudes is considered to be smooth, and the image distortion from the assumption that radiation at latitudes of $\pm 2.5^{\circ}$ to be 0 is negligible. Even after this baseline subtraction, scanning noise in the latitude direction remained in images due to a fast fluctuation of the system gain and the atmospheric absorption. The PRESS method (Sofue \& Reich 1979) was applied to reduce this scan noise. Several spurious remained in an image were replaced by interpolation of the adjacent data. After these processing, the images of two polarization on the two days were averaged, then were smoothed with a circular Gaussian function with a FHWM of $0.5'$ (synthesized beam width $= 4.3'$). The noise level of the final map is $1 \sigma = 10$ mK. Residual scanning noise in the latitude direction remains across the map, but does not exceed 20 mK. Smoothly spread galactic emissions overlapped in the image in addition to the emission from the galactic center. The smoothly spread galactic emission was removed, using the background emission filtering method (Sofue \& Reich 1979).

\subsection{Radio recombination line observation}
The recombination line of H92$\alpha$ (8309.382 MHz) was observed in a period from 2011 to 2014. Since the frequency of the line is covered by the band for our continuum observations, the parameters of the beam are the same as that of the continuum observations. The observed area was $-1.00^{\circ} \le l \le 0.53^{\circ}$ and $0.10^{\circ} \le b \le 1.25^{\circ}$, covering the entire GCL. The sampling was made at a grid of $24 \times 24$ points with intervals of $4'$ and $3'$ in longitude and latitude, respectively.
Pointings at $l = -60'$ and $-56'$ within $0.60^{\circ} \le b \le 0.75^{\circ}$ and pointings at $l = 24'$, $28'$, and $32'$ within $0.60^{\circ} \le b \le 1.25^{\circ}$ were not observed.
The total number of the observation points was 526. Although the sampling is coarser than the Nyquist sampling, this is the highest angular-resolution mapping observation with the recombination line of the entire GCL achieved so far. The position switching method was used for the observations with the OFF point at $l = 0.50^{\circ}$ and $b = 2.00^{\circ}$ (RA = 17h 39m 06.0s, Dec = $-27^{\circ}$ $27'$ $28"$, J2000.0). The observation was made simultaneously with both the right and left circular polarizations. The observing band width was 32 MHz during an early period of observation and 8 MHz during a late period, which encompasses H92$\alpha$. The number of frequency channel was 512 or 128, and consequently the band width of one channel was $\Delta \nu = 62.5$ kHz, corresponding to 2.25 km s$^{-1}$ in velocity. The system noise temperature was typically 45 to 50 K, and 70 K in bad weather. The system noise temperature was measured using a noise source or by using thermal radiation from the nearby mountains as black body radiation. The noise temperatures of the noise source and mountains were calibrated with the room temperature absorber. The integration time of each observation point was 45 minutes, and the observation duration of each point was 2 hours including the antenna slew time and the OFF point observation. After averaging the data of two polarizations, the rms noise per channel of the average spectra was $1 \sigma = 3.5$ mK. The observable time of the galactic center from Yamaguchi is about 6 hours per day, and accordingly 3 points were observed per day. The total observation time was about 1100 hours.

To obtain the line parameters, baseline fitting with a 3rd order polynomial was applied to the data in the velocity range from $-100$ to $+100$ km s$^{-1}$. If a line was visually detected, Gaussian function fitting was performed to obtain the antenna temperature, the LSR velocity, and the line width of the line. The detection threshold of the line peak temperature was set to $2\sigma$.

\subsection{Observations at 23 GHz} 
Observations for H65$\alpha$ of the GCL were also made with the Tomakomai 11 m radio telescope (Sorai et al. 2008). The observation position was $l = -0.60^{\circ}$ and $b = 0.45^{\circ}$ where H92$\alpha$ had been strongly detected with the Yamaguchi 32 m. The rest frequency of H65$\alpha$ is 23404.28 MHz and the beam size of Tomakomai 11 m at this frequency was $4.2'$, which is nearly the same size as that of the observation of H92$\alpha$ with the Yamaguchi 32 m. The system noise temperature was 150 K, the main beam efficiency of the antenna was $46 \%$, and the polarization was the LHCP. The position switching observation was used. The observations were carried out from 2014 November to 2014 December for 15 days, and the total on-source integration time was 6.51 h.

As a result, a weak line was detected. The antenna temperature $T_{\rm L} = 10.0 \pm 2.7$ mK, $V_{\rm LSR} = -1.74 \pm 2.77$ km s$^{-1}$, and the line width $\Delta V_{\rm D} = 11.92 \pm 4.79$ km s$^{-1}$ were obtained by Gaussian fitting. This is consistent with the result of our H92$\alpha$ observations with the Yamaguchi 32 m in intensity and LSR velocity. However, the antenna temperature did not exceed the detection threshold adopted in this paper of $2 \sigma$ ($1 \sigma = 6.7$ mK). Therefore, this result is not discussed in the following sections.

\section{Results}
\subsection{Spatial intensity distribution of radio continuum}
Figure 1a shows the central $3^{\circ} \times 3^{\circ}$ of the 8 GHz continuum map. Strong emission of Sgr A and the radio arc at the center are prominent. Sgr B2 at a longitude of $0.6^{\circ}$ and Sgr C at $-0.6^{\circ}$ are also remarkable. The GCL extending from the galactic plane to the north side is clearly visible. Figure 1b shows the expanded image of the GCL. The ridges of the GCL on the east side (left) and the west side (right) are connected at a latitude of $1.2^{\circ}$ and longitude of $-0.3^{\circ}$ to form a dome-like structure. These structures show a good morphological agreement with that observed by the Effelsberg 11 cm survey (Reich et al. 1990b, FWHM $= 4.3'$).

\begin{figure}
 \begin{center}
  \includegraphics[width=16cm]{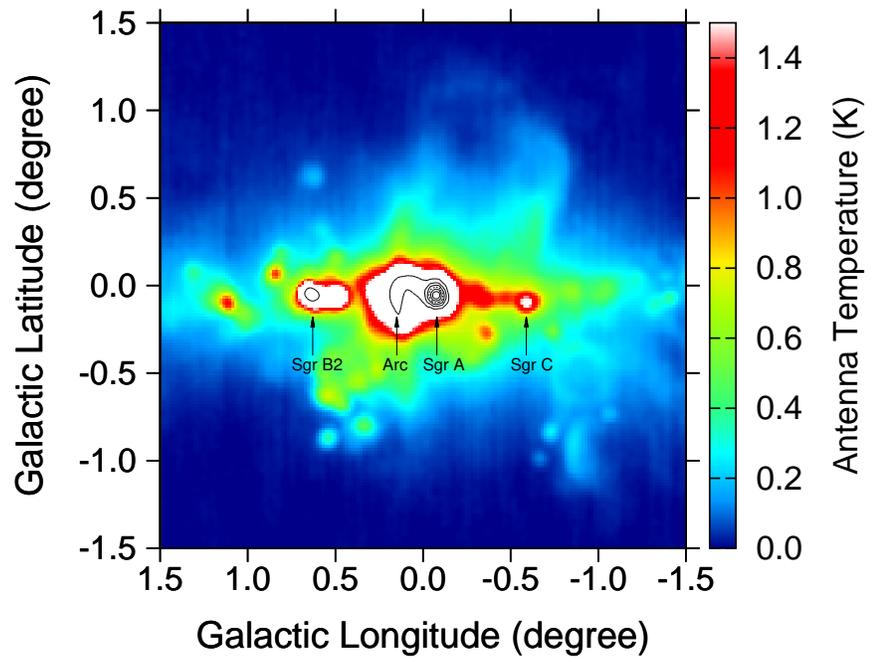}
  \includegraphics[width=16cm]{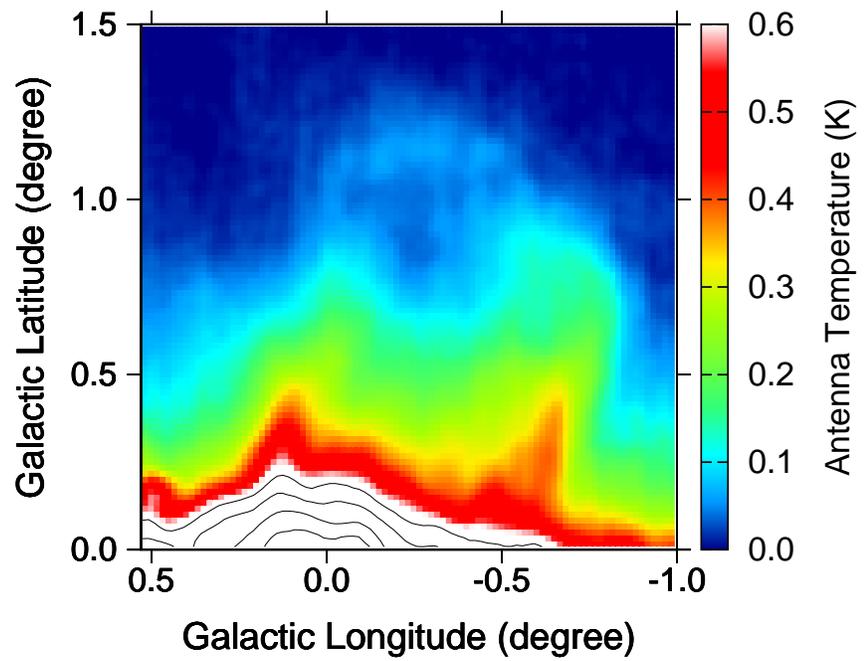}
 \end{center}
\caption{1a. (upper panel) Antenna temperature (intensity) distribution of 8 GHz radio continuum. 1b (lower panel) close-up view of the GCL.}
\label{fig:fig1}
\end{figure}

\subsection{Spatial intensity distribution of H92$\alpha$}
Figure 2 shows a line spectrum obtained at $l = -0.60^{\circ}$ and $b = 0.40^{\circ}$ as an example, where the derived parameters are the peak antenna temperature $T_{\rm L} = 68 \pm 3$ mK (highest of all data), the LSR velocity $V_{\rm LSR} = -2.03 \pm 2.29$ km s$^{-1}$, and the line width (FWHM) $\Delta V_{\rm D} = 13.45 \pm 4.53$ km s$^{-1}$.

\begin{figure}
 \begin{center}
  \includegraphics[width=16cm]{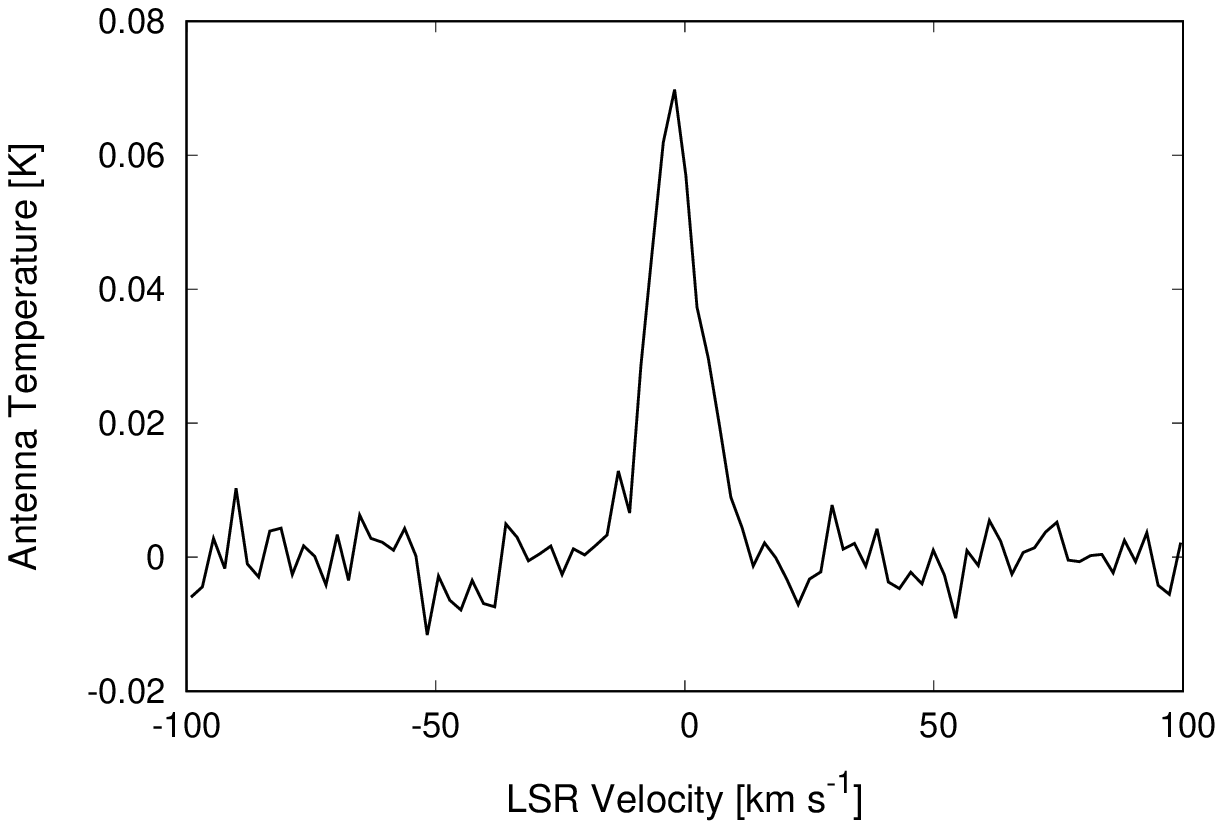} 
 \end{center}
\caption{Radio recombination line H92$\alpha$ at $l = -0.60^{\circ}$ and $b = 0.40^{\circ}$.}\label{fig:fig2}
\end{figure}

Figure 3 shows the peak line temperature (hereafter, line intensity) distribution of H92$\alpha$. As with the intensity distribution of the continuum in figure 1b, two ridges at eastern and western sides extending toward the north from the galactic plane are clearly visible. In the west ridge at a latitude of $b = 0.10^{\circ}$, a line intensity peak is at around $l = -0.45^{\circ}$. The peak position is slightly offset to the west as it extends to the north from the galactic plane, and at $b = 0.85^{\circ}$ the peak is at $l = -0.7^{\circ}$. From this point, the ridge changes the angle of the elongation and extends toward the northeast. The maximum intensity in the west ridge is at $l = -0.60^{\circ}$ and $b = 0.40^{\circ}$, where the peak line temperature reaches 60 mK (figure 2). For the east ridge, it extends from ($l, b$) = ($-0.07^{\circ}, 0.10^{\circ}$) to ($0^{\circ}, 1.2^{\circ}$). The peak line temperature of the east ridge reaches a maximum of 40 mK at $l = 0.00^{\circ}$ and $b = 0.20^{\circ}$. The two ridges appear to connect to each other at around $b = 1.2^{\circ}$ and form a loop-like structure overall, in the sense that the line intensity is much lower inside the loop. To summarize these results, the recombination line distribution of the GCL extends from the galactic plane to the north in the east and west ridges, forming a loop structure from $b = 0.1^{\circ}$ to $1.25^{\circ}$ in latitude. The maximum longitudinal angular extent of the structure is $0.9^{\circ}$ at $b = 0.8^{\circ}$. This is in good agreement with the radio recombination-line observations with the HCRO 25 m (Law et al. 2009) and with the Parkes 64 m (Alves et al. 2015). The intensity distributions of the continuum and radio recombination line of the GCL are in good agreement especially at high latitudes.

\begin{figure}
 \begin{center}
  \includegraphics[width=16cm]{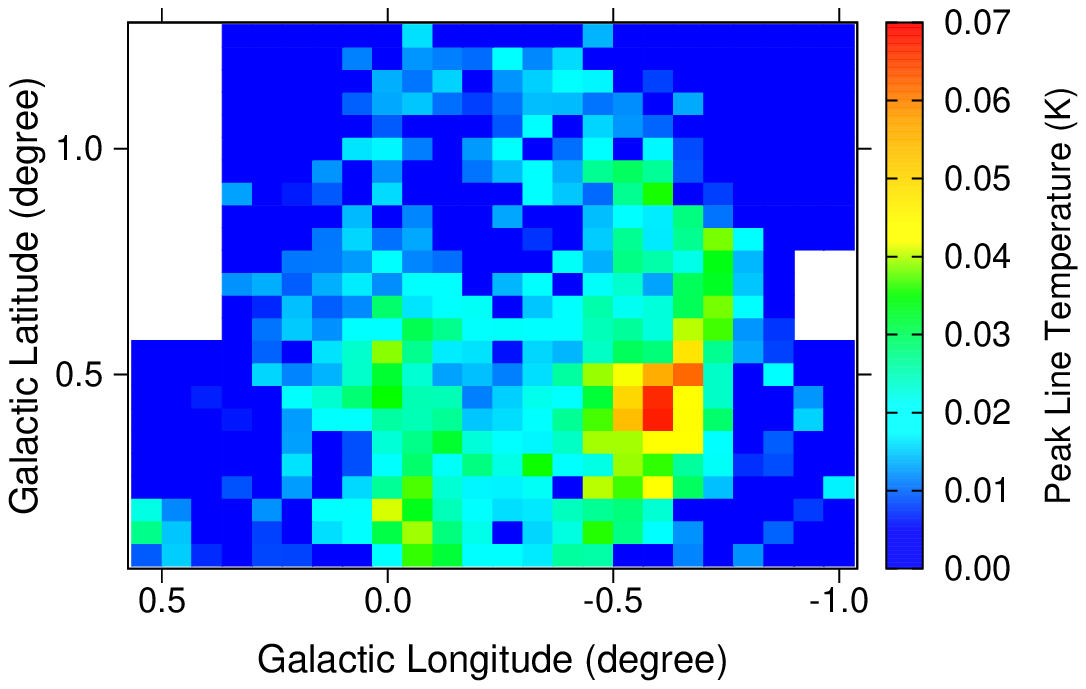} 
 \end{center}
\caption{Peak line temperature (intensity) distribution of the radio recombination line H92$\alpha$.}\label{fig:fig3}
\end{figure}

\begin{figure}
 \begin{center}
  \includegraphics[width=16cm]{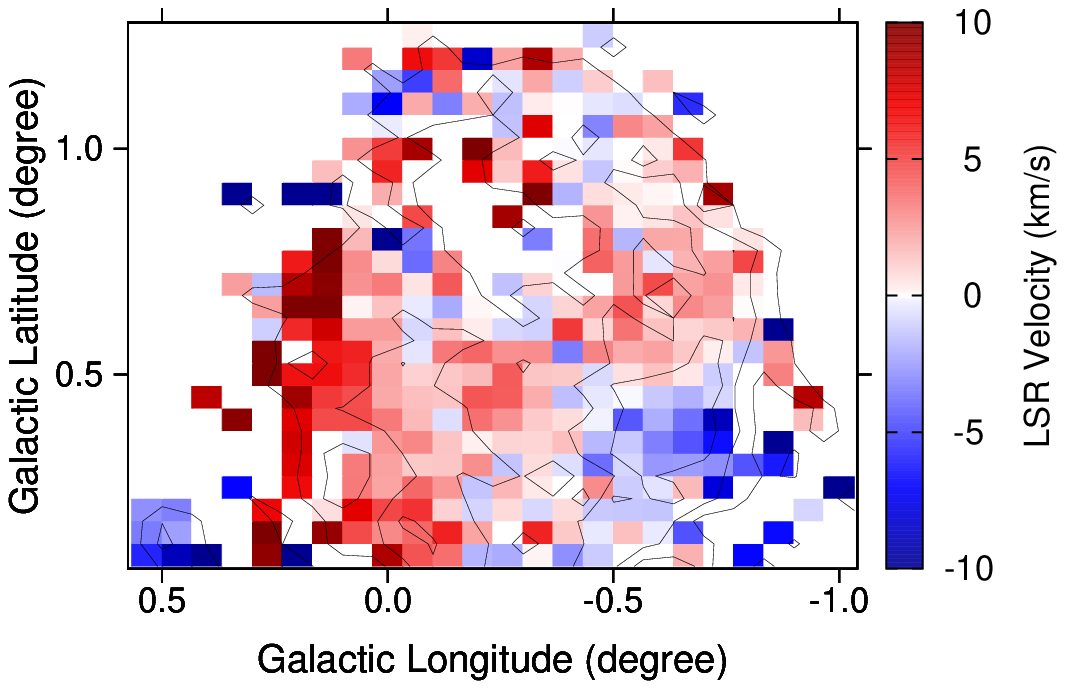} 
 \end{center}
\caption{LSR velocity distribution of the radio recombination line H92$\alpha$ with contours of line peak temperature at 0.008, 0.023, and 0.038 K.}\label{fig:fig4}
\end{figure}

\subsection{Velocity distribution}
Figure 4 shows the radial velocity distribution of the radio recombination line. In the east ridge the velocity is positive, ranging up to $+10$ km s$^{-1}$, whereas in the west ridge, the velocity is negative at low latitudes of $b < 0.5^{\circ}$ and shows the lowest velocity of $-4$ km s$^{-1}$. At $b = 0.45^{\circ}$, a velocity gradient of $7.5$ km s$^{-1}$ deg$^{-1}$ from the west to east is observed, which is consistent with that observed by Law et al. (2009). However, at $b = 0.5^{\circ}$ and higher, no clear velocity gradient is observed. Almost all the points observed in the GCL show small velocities of around 0 km s$^{-1}$, which are much smaller than the galactic rotation velocity. In addition to the spatial distribution of the radio recombination line intensity, the line velocity also shows a continuous distribution over the entire GCL. Therefore, the thermal plasma identified with the radio recombination line is a single continuous object as discussed by Law et al. (2009) and Alves et al. (2015).

\subsection{Line width and electron temperature}
The electron temperature $T_{\rm e}$ of the plasma forming the GCL can be estimated from the line width (FWHM) $\Delta V_{\rm D}$ of the radio recombination line with the equation,
\[
T_{\rm e} = \frac{m_{\rm p} \Delta V_{\rm D}^{2}}{8 k_{\rm B} \log 2},
\]
where $m_{\rm p}$ is the mass of a proton and $k_{\rm B}$ is the Boltzmann constant. The typical line width is estimated from our data to be $\Delta V_{\rm D} = 14.13 \pm 1.45$ km s$^{-1}$ as the averaged value of those at four points where the line intensity is particularly high ($T_{\rm L} > 55$ mK). Substituting it to the equation yields the electron temperature $T_{\rm e} = 4360 \pm 900$ K. This temperature is an upper-limit because the effect of gas turbulent motion that expands the line width as well as thermal broadening is ignored. The electron temperature of a typical HII region ranges from 5000 to 10000 K, whereas our result shows a slightly lower temperature than it. Law et al. (2009) derived the electron temperature to be $3960 \pm 120$ K from the line width of $13.5 \pm 0.2$ km s$^{-1}$, which agrees with our result. The result by Alves et al. (2015) also showed a narrow line width of $12 \pm 1$ km s$^{-1}$. The electron temperature of $T_{\rm e} = 4360$ K is adopted for the thermal plasma of the GCL in the following sections of this paper.

\section{Discussion}
\subsection{Correlation between the continuum and the radio recombination line}
The peak line temperature of a radio recombination line is proportional to that of free--free emission from the thermal plasma. Figure 5 shows a scatter plot of the observed peak line temperatures of the radio recombination line and the antenna temperature of the continuum of the GCL. The data points at low latitudes of $b < 0.7^{\circ}$ (black points) are scattered toward right in figure 5, presumably because the contamination to the radio continuum from the synchrotron radiation of the unrelated radio sources in the line of sight is more significant at low latitudes. On the contrary, the peak line temperature of the radio recombination line is found to be proportional to that of the continuum at higher latitudes of $b \ge 0.7^{\circ}$ (red points). This result suggests that the continuum emission of the GCL observed at the high latitude is dominated by free-free emission from thermal plasma as has been discussed by Law et al. (2009). Seven points with high line temperature ($T_{\rm L} > 45$ mK) are visible in the upper part of figure 5. These are observed around $b = 0.45^{\circ}$ of the west ridge at which the radio recombination line is strong. We obtain the temperature ratio of the line ($T_{\rm L}$) to continuum ($T_{\rm c}$) to be $T_{\rm L} / T_{\rm c} = 0.19 \pm 0.01$ from the data of the high latitude and the seven points for the strongest line emission. The theoretical line-to-continuum ratio is, assuming the local thermal equilibrium (LTE, Rohlfs \& Wilson 2000), given by:
\[
\frac{T_{\rm L}}{T_{\rm c}} \left(\frac{\Delta v}{\rm km~s^{-1}}\right)=\frac{6.985 \times 10^{3}}{a(\nu, T_{\rm e})} \left(\frac{\nu}{\rm GHz}\right)^{1.1} \left(\frac{T_{\rm e}}{\rm K}\right)^{-1.15} \frac{1}{1+N(\rm He^{+})/N(\rm H^{+})}.
\]
Adopting $\Delta v = 14.1$~km~s$^{-1}$, $T_{\rm e} = 4360$~K, $a(\nu, T_{\rm e}) = 1$, $\nu = 8.309$~GHz, and $N(\rm He^{+})/N(\rm H^{+})=0.08$, $T_{\rm L} / T_{\rm c} = 0.31$ is obtained, which is 1.6 times higher than the observed value. By comparing the line-to-continuum ratio $T_{\rm L} / T_{\rm c} = 0.19$ obtained for the data of $b \ge 0.7^{\circ}$ and the seven points with high line temperature and the theoretical value $T_{\rm L} / T_{\rm c} = 0.31$, it is derived that about 60\% of the continuum emission is thermal and about 40\% is non-thermal in the region of $b \ge 0.7^{\circ}$. This fraction varies with $b$. The observed $T_{\rm L} / T_{\rm c}$ is as low as 0.14 in the area of $b \ge 0.5^{\circ}$, while it is 0.28 for $b \ge 1.0^{\circ}$, where the thermal emission reaches 90\% of the continuum.

\begin{figure}
 \begin{center}
  \includegraphics[width=16cm]{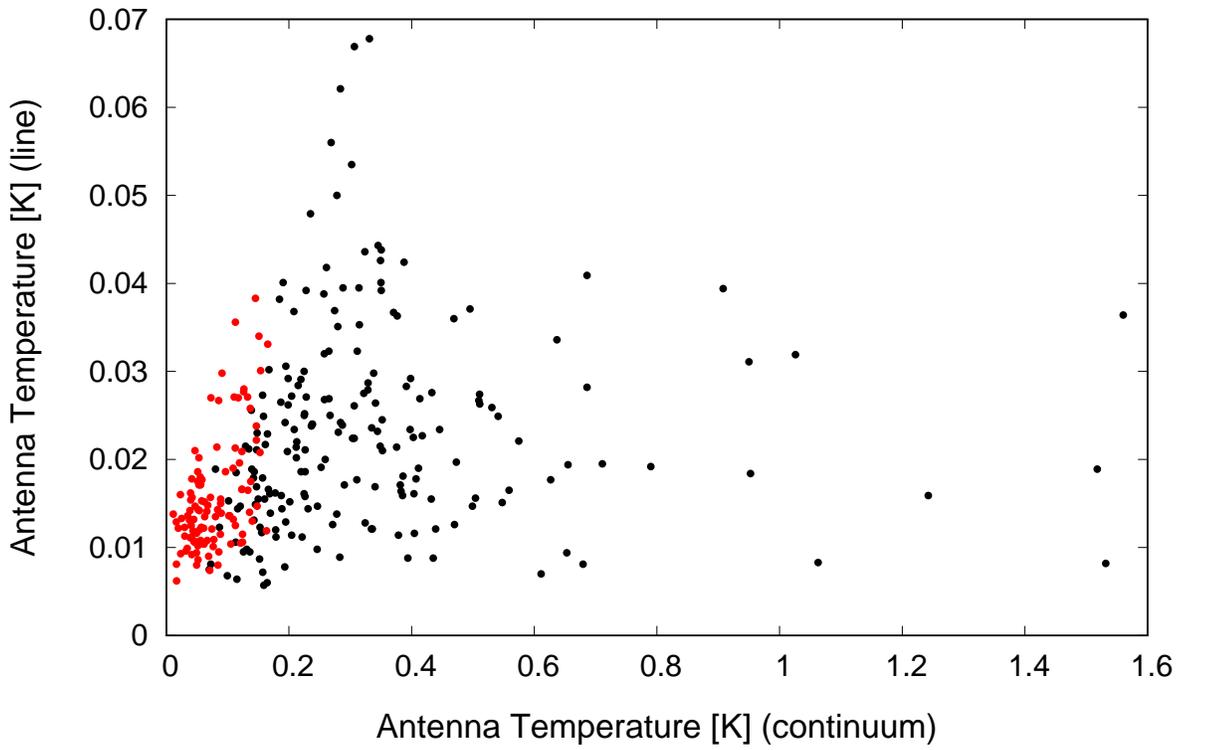} 
 \end{center}
\caption{Intensity distribution of the observed radio recombination line and continuum of the GCL. Black and red points indicate the data at $b < 0.7^{\circ}$ and $b \ge 0.7^{\circ}$, respectively.}
\label{fig:fig5}
\end{figure}

\subsection{Are the Radio Arc and/or Sgr C related to the GCL?}
As shown in figure 5, most of the continuum emission observed at lower latitudes might be unrelated to the GCL. In this section, we discuss the spatial structure of the GCL and potential relation or no relation with other major objects in the region observed by radio continuum. The continuum intensity is noticeably large at or near the radio arc. We discuss first that if the radio arc is related to the GCL. Figures 6 (a) - (d) show the longitudinal profiles of the antenna temperatures of the continuum and radio recombination lines at four latitudes of $b = 0.10^{\circ}$ to $0.25^{\circ}$. At a latitude $b = 0.10^{\circ}$ (Figure 6a), the antenna temperature of the continuum peaks at $l = 0.11^{\circ}$, which is in the north side of the radio arc or coincides with the southern edge of the polarization plume (Tsuboi et al. 1986), where the antenna temperature of the continuum is 2.7 K. The peak line temperature distribution of the radio recombination line shows two peaks corresponding to the east and west ridges of the GCL. The eastern peak does not coincide with the radio arc, but is close to another continuum peak near $l = -0.07^{\circ}$. The line temperature of the radio recombination line at the peak position is 36 mK, whereas that at the radio arc is below the detection limit of 8 mK. Similarly, at latitudes of $b = 0.15^{\circ}$ to $0.25^{\circ}$, the eastern peaks of the radio recombination line are not at the ridge of the highest continuum peaks, which are extended north from the radio arc but are accompanied by the peaks of the continuum at around $l = 0^{\circ}$.

\begin{figure}
 \begin{center}
  \includegraphics[width=9cm]{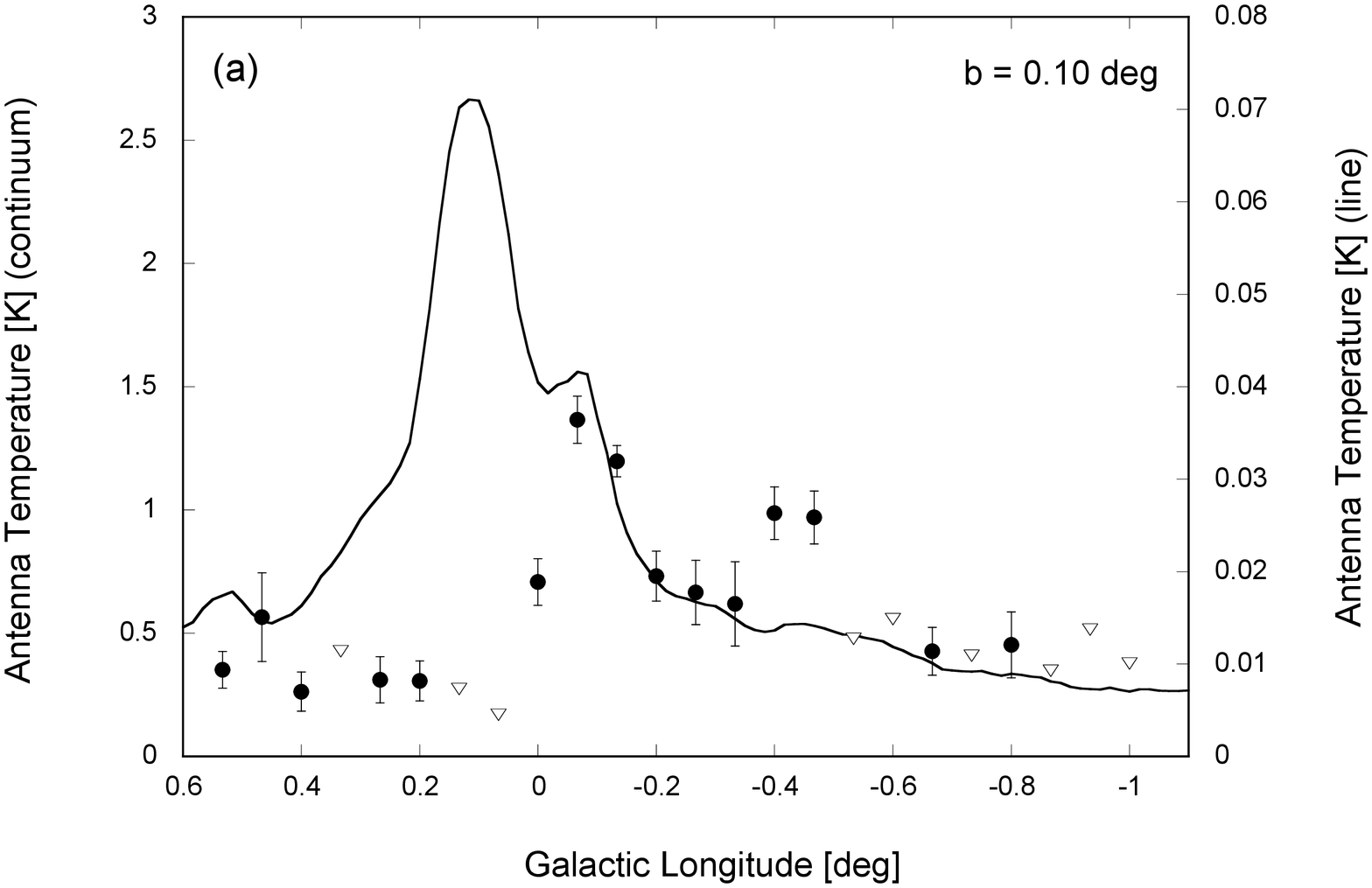} 
  \includegraphics[width=9cm]{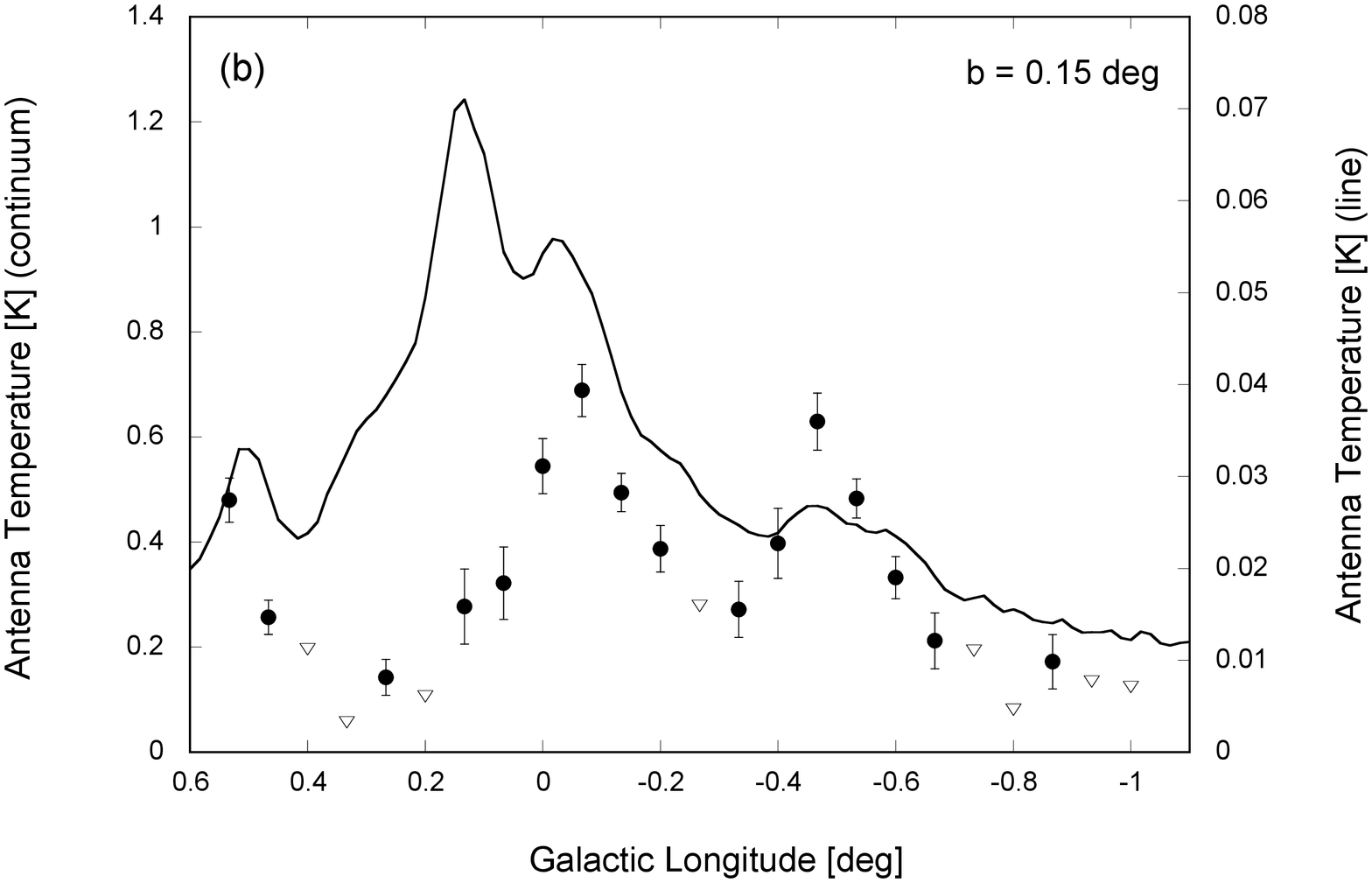} 
  \includegraphics[width=9cm]{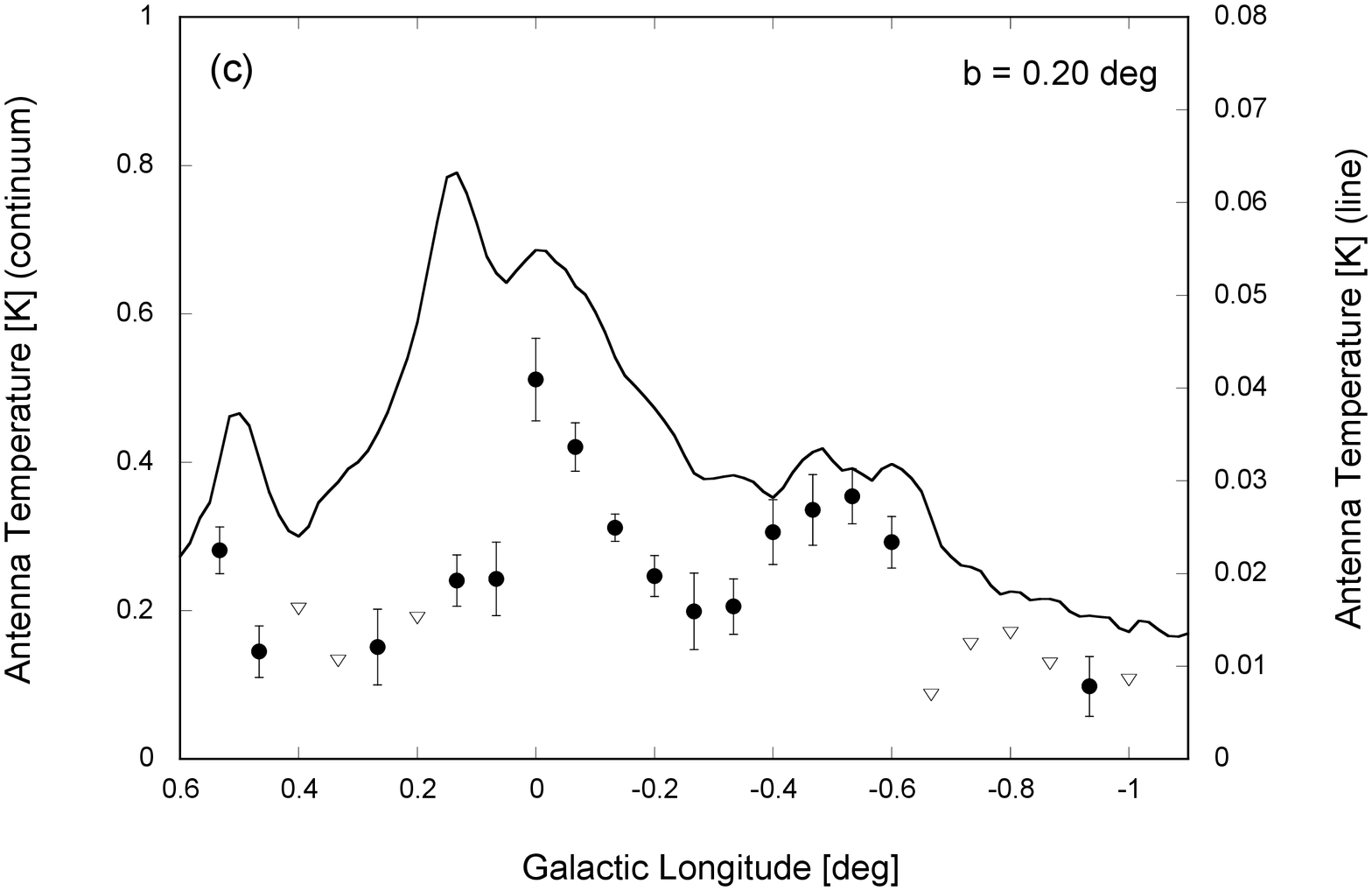} 
  \includegraphics[width=9cm]{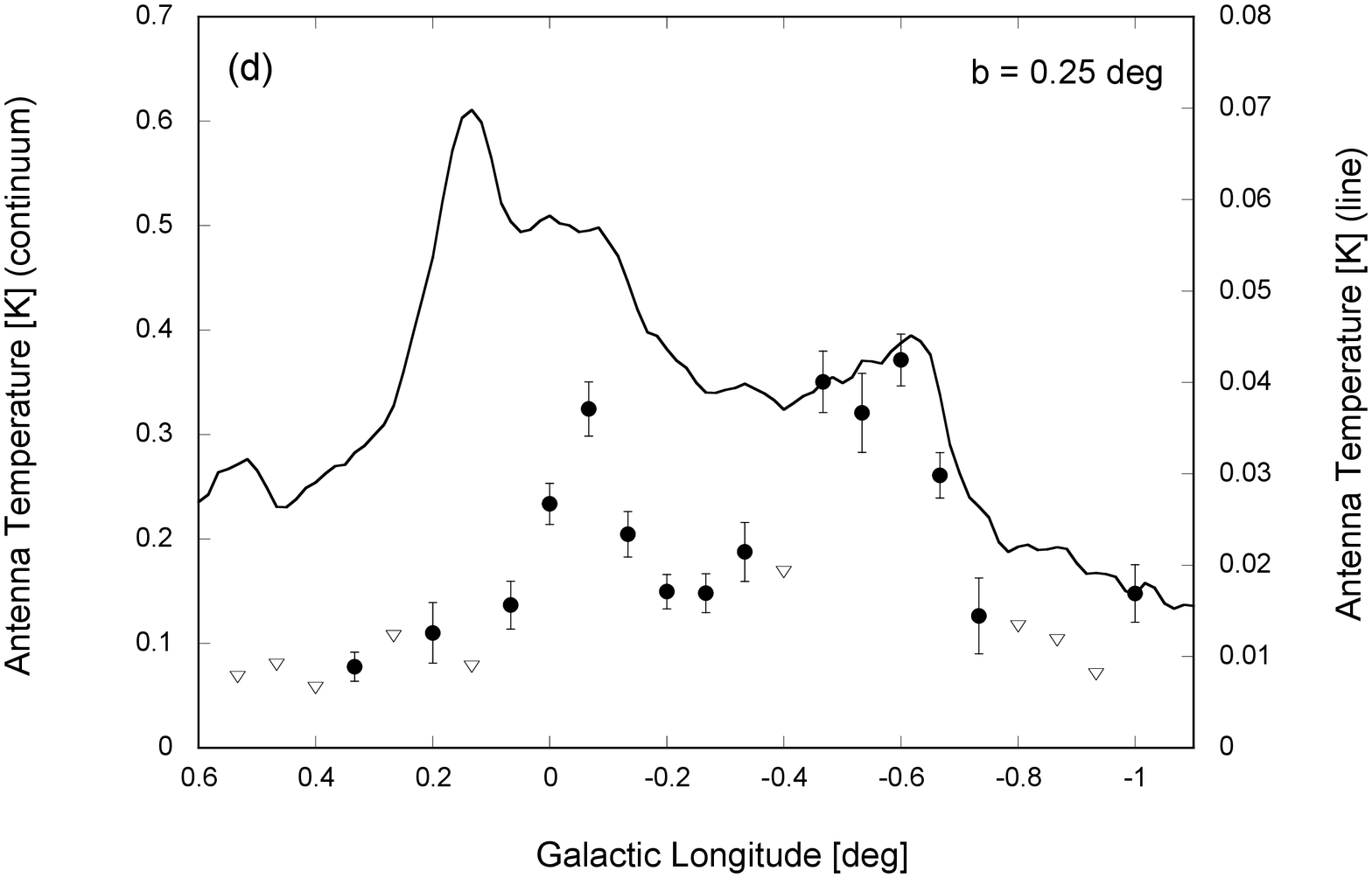} 
 \end{center}
\caption{Longitudinal profiles of intensity distribution of the continuum and radio recombination line at four different galactic latitudes. Solid lines show the antenna temperature of the continuum (left vertical axis), filled circles and open triangles show that of the radio recombination line (right vertical axis) and its upper limit, respectively. The error bar of the the radio recombination line is obtained by Gaussian fitting.  (a) latitude $b = 0.10^{\circ}$, (b) $0.15^{\circ}$, (c) $0.20^{\circ}$, (d) $0.25^{\circ}$.}
\label{fig:fig6}
\end{figure}

To investigate the relationship in detail between the peak positions of the continuum and radio recombination line for the whole regions of the GCL, we fit the distributions of the antenna temperature at each latitude with a multiple-component Gaussian function, and determine the peak positions of the intensity. The number of Gaussian components for the radio recombination line is varied, depending on the latitude: 2 (for $b = 0.10$ to $1.10^{\circ}$), 1 ($b = 1.15, 1.20^{\circ}$), and 0 ($b = 1.25^{\circ}$). The continuum data are fitted with 1 to 5 Gaussian components. Figure 7 shows the derived peak positions of each Gaussian component. The global structure of the GCL with the two distinctive ridge-like features in the east and west is broadly reproduced with both the continuum and recombination-line peak positions. We find that the west ridge in the radio recombination line runs along the two (west) ridges of the continuum, and that the distributions of the two emissions agree with each other reasonably. In contrast, the east ridge in the radio recombination line does not coincide with the northern part of the radio arc. The position of the radio arc is offset by $0.2^{\circ}$ from the ridge of the radio recombination line. This offset is sufficiently larger than the fitting error of the peak position of the radio recombination lines. The separation of $0.2^{\circ}$ corresponds to 30 pc at the center of the galaxy. In the ridge on the west side, the distribution of the west most continuum ridge and the ridge of the radio recombination line show a layered distribution with $0.08^{\circ}$ (13 pc) separation as discussed by Law (2010). The separation of the radio arc and the radio recombination line ridge on the east side is more than twice the separation of the west ridges.

\begin{figure}
 \begin{center}
  \includegraphics[width=16cm]{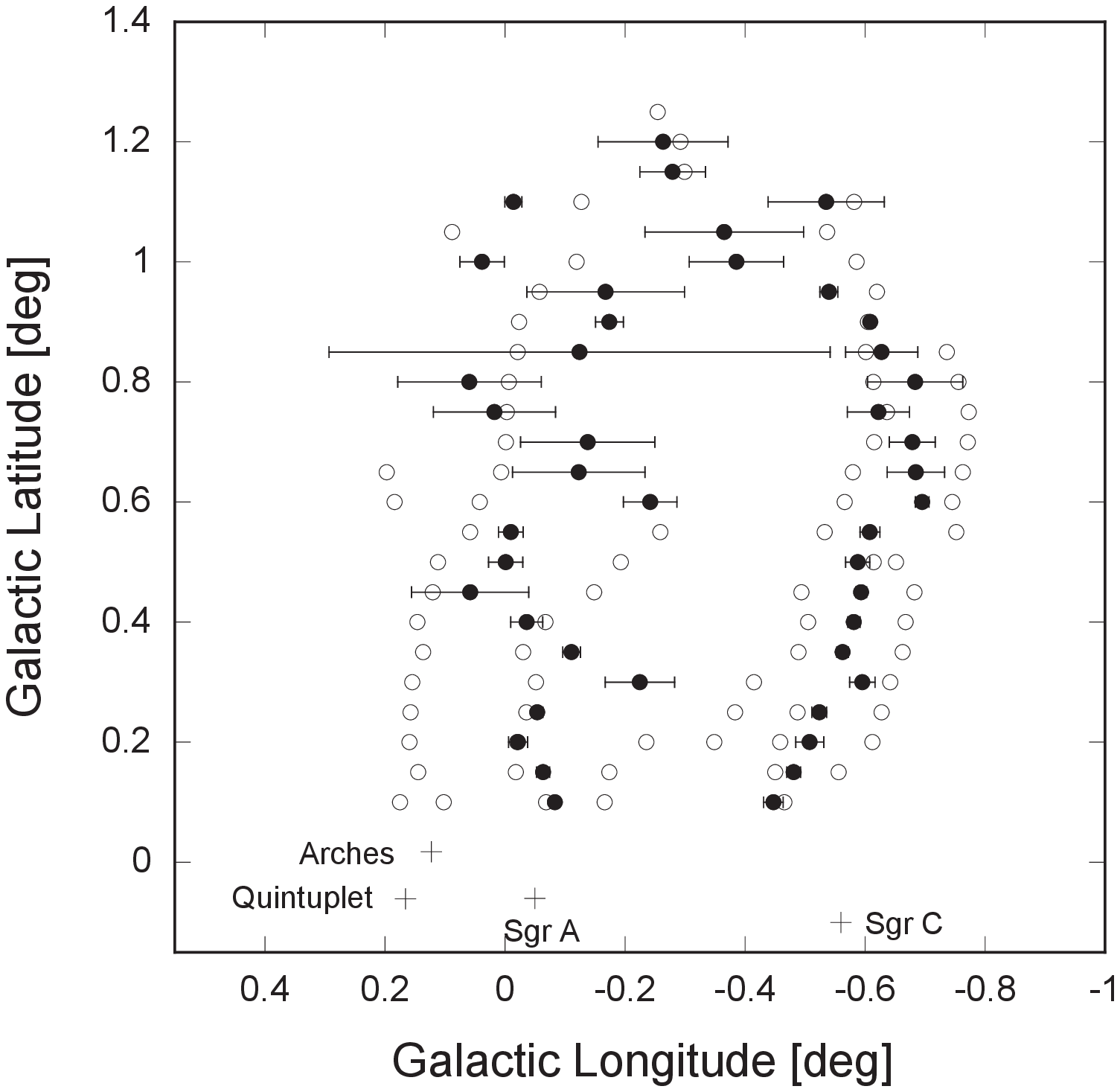} 
 \end{center}
\caption{Distribution of the peak positions. Open and filled circles indicate peak positions of the continuum and radio recombination lines, respectively. The error bar of the peak position of radio recombination line shows the fitting error. The positions of the large star clusters (Quintuplet, Arches), Sgr A, and Sgr C are indicated by crosses.}
\label{fig:fig7}
\end{figure}

The ridge in the continuum emission, which runs at around $l = 0.0^{\circ}$, is clearly visible from $b = 0.1^{\circ}$ to $0.4^{\circ}$ in figure 7. This ridge is also seen in the image of 10 GHz by the Nobeyama 45 m (figure 8 in Sofue 1985) and the image of the 8 GHz Parkes 64 m (Haynes et al. 1992). This $l = 0.0^{\circ}$ ridge extends substantially north to $b = 1.0^{\circ}$, and shows an overlapped distribution with the ridge of radio recombination line. Thus, a large fraction of the radiation at the $l = 0.0^{\circ}$ ridge is the free - free radiation of the thermal plasma of the GCL, and this continuum ridge is considered to be the counterpart of the ridge of the radio recombination line. The peak position of the ridge is obscured at around $b = 0.4^{\circ} - 0.6^{\circ}$. This may be because the $l = 0.0^{\circ}$ ridge apparently overlaps the continuum ridge extended from the radio arc.

In the galactic center image at 74 MHz with the VLA (Brogan et al. 2002), the GCL is visible as an absorbing structure. A plausible interpretation is that high-brightness temperature synchrotron radiation from sources, which perhaps are situated near the galactic center, is absorbed by the thermal plasma of the GCL in the line of sight. The east ridge of the absorption structure is not located in or near the radio arc but runs along the ridge at around $l = 0.0^{\circ}$. Yusef-Zadeh et al. (2009) indicates, in the mid-infrared image, a ridge extending from the galactic center ($l \sim 0^{\circ}$) to the north side as a potential counterpart of the east ridge of the radio GCL. Note that it is not at the position of the radio arc but at the position of the continuum ridge at $l = 0.0^{\circ}$. In the polarization distribution image of Tsuboi et al. (1986), there is a region where the degree of polarization is low (Pol $< 4 \%$) at $l \sim 0^{\circ}$ and $b = 0.3$ to $0.5^{\circ}$. This region coincides with the position of the $l = 0.0^{\circ}$ ridge, also with the east ridge of the GCL observed in the radio recombination line. These properties of the $l = 0.0^{\circ}$ ridge are consistent with that the continuum emission in the $l = 0.0^{\circ}$ ridge is dominated by free-free radiation from thermal plasma in the GCL.

Sofue (1985) argued that the west side of the GCL is connected to Sgr C, and similar arguments have been made in subsequent papers (e.g., Law 2010). However, the peak distribution of the radio recombination line shown in figure 7 deviates from Sgr C ($l = - 0.56^{\circ}, b = -0.10^{\circ}$) and points towards the larger longitude as it becomes closer to the galactic plane, and reaches near $l = -0.4^{\circ}$ at $b = 0.1^{\circ}$. The ridge of the radio recombination line and Sgr C are separated by $0.2^{\circ}$ (30 pc). Similar to the radio recombination line, an infrared ridge in the image of Spitzer / IRAC (Arendt et al. 2008; Law 2010) which would relate to the west ridge of the GCL seems not to connect to Sgr C. Although it cannot be concluded, these ridges of the GCL would be unrelated to Sgr C.

Since the observational results of radio recombination line of the GCL suggest that the GCL would not physically connect to either the radio arc or Sgr C, which exist in the central region of the galaxy, the GCL may not be situated physically in the galactic center region. 

\subsection{Velocity and Location of the GCL}
Gas structures existing at the center of the galaxy, in general, are known to move at high speed ($> 100$ km s$^{-1}$). For example, Sofue (1985, 1996) reported a high velocity of CO molecular clouds from $-150$ to $+150$ km s$^{-1}$ in the region that overlaps with the GCL. Fukui et al. (2006) found the molecular loops I and II by CO observations near the galactic center, which have similar characteristics and apparent size to those of the GCL ($\sim 1^{\circ}$ or 150 pc at 8.5 kpc), and found that they have large central velocities of $-95$ km s$^{-1}$ and $-70$ km s$^{-1}$ and the velocity ranges of 80 km s$^{-1}$ and 60 km s$^{-1}$, respectively. Compared with the parameters for the loops I and II, the central velocity and amount of the velocity range of the GCL of +3 km s$^{-1}$ and 14 km s$^{-1}$, respectively, are much smaller. 

Law et al. (2009) explained the low velocity of the ionized gas in the GCL by the gas dynamics in a barred gravitational potential (Binney et al. 1991). However, this argument is applicable only to objects with small spatial sizes. The size of the GCL exceeds 100 pc if it exists in the galactic center. If such a large object exists in the center of the galaxy, the velocity structure of the object would be inevitably affected by the galactic rotation like molecular loops I and II in contrast to that the observed radial velocity in the GCL that falls in the range of $-4$ to $+10$ km s$^{-1}$. These facts suggest that the GCL is not in the galactic center region. The longitude-velocity diagram of CO emission by Dame et al. (2001) shows that a wide range of radial velocities are observed from the galactic center, including $0$ km s$^{-1}$. Such low-velocity gases are observed in the direction of the center of the galaxy, but they are actually located away from the center of the galaxy.

The velocity gradient inside the GCL follows the same trend as with the galaxy rotation, i.e., the velocity is positive on the east side and negative on the west side. The observed velocity gradient is 8.95 km s$^{-1}$ deg$^{-1}$ at a latitude $b = 0.3^{\circ}$. On the assumption that this velocity gradient originates in the galactic rotation, the distance between the galactic center and the GCL is estimated as follows. Suppose that the distance between the solar system and the galactic center is 8.5 kpc, that the rotation velocity of the galaxy is constant at 220 km s$^{-1}$, and that the GCL is at $l = 0^{\circ}$ and rotating in line with the galactic rotation. Then, the distance from the galactic center to the GCL is derived to be 2.6 kpc to reproduce the velocity gradient of 8.95 km s$^{-1}$ deg$^{-1}$ in the GCL by the galactic rotation (equation (5) in Sofue 2006), although this estimate has a significant uncertainty.

On the other hand, three pieces of evidence suggest that the GCL is not a local object close to the solar system as follows. (A) The strong absorption of H$\alpha$ radiation near the galactic plane (Gaustad et al. 2001) indicates that the GCL is far from the solar system. (B) The radio recombination line intensity is small at the upper part of the GCL in contrast to the lower part. This suggests that the GCL is an object with a scale of spatial extent comparable with the thickness of the galactic plane, 35 pc for molecular gas and and 100 pc for HI gas (Malhotra 1994, 1995). (C) Dependency on latitude is observed in the velocity gradient of the radial velocity inside the GCL. Whereas the gradient is clearly identified at positions near the galactic plane, it is unclear at higher galactic latitudes. This suggests that the GCL is a large-scale object, as is the case with (B). Therefore, the estimated distance of 2.6 kpc from the galactic center to the GCL is reasonable.

We point out that an observational result has been reported that favors the interpretation that the GCL is physically situated in the galactic center region. Law et al. (2011) showed that the magnetic field parallel to the line-of-sight direction is well aligned in the GCL by the Faraday rotation observation, and that the global magnetic-field direction is reversed at $l = -0.3^{\circ}$. This longitude corresponds to almost the center of the GCL observed with the radio recombination line. Given that the Faraday rotation is caused via interaction between magnetic field and thermal plasma, the results by Law et al. (2011) suggest that the GCL contains a global toroidal magnetic field. The straightforward interpretation of this global toroidal magnetic-field structure is that the GCL is physically in the central region of the galaxy. If the GCL is located in the forground of the galactic center region, as the other pieces of evidence described in this subsection suggest, we should suppose that a toroidal magnetic field accidentally exists at the location of the GCL.

\subsection{Physical parameters and formation mechanism of the GCL}
In the following discussion, we assume the distance of 6 kpc from the solar system (2.5 kpc from the galactic center) to the GCL. At this distance, the real size of the GCL is 90 pc and 130 pc in longitude and latitude, respectively. Emission Measure ($EM$) can be estimated from the antenna temperature of the radio recombination line as follows. First, the optical depth $\tau_{\rm c}$ and $EM$ of the radio continuum (free - free emission) are related by Oster (1961) with the formula:
\[
\frac{EM}{\rm cm^{-6}pc}=\frac{\tau_{\rm c}}{3.014 \times 10^{-2} T_{\rm e}^{-1.5} \nu^{-2}(\log(4.995 \times 10^{-2} \nu^{-1})+1.5\log T_{\rm e})},
\]
where $\nu$ is the observation frequency 8.309 GHz, $\tau_{\rm c} = (T_{\rm c}/\eta_{\rm B}) / T_{\rm e}$ is an optical depth, $\eta_{\rm B} = 0.70$ is a beam efficiency, $T_{\rm e} = 4360$ K is the electron temperature. The antenna temperature $T_{\rm c}$ of the continuum is estimated from the antenna temperature $T_{\rm L}$ of the radio recombination line, using the line-to-continuum ratio $r = T_{\rm L} / T_{\rm c} = 0.31$. The $EM$ ranging from $5.8 \times 10^{2}$ to $6.4 \times 10^{3} \rm cm^{-6}pc$ is estimated from these observed quantities. The electron density $n_{\rm e}$ is estimated with the $EM$ by,
\[
\frac{n_{\rm e}}{\rm cm^{-3}} \cong \sqrt{\frac{EM}{L}},
\]
where $L$ (pc) is a depth of the plasma along the line of sight. This path length was derived by Law (2009) as 50~pc assuming a shell geometry for the GCL and the distance of 8~kpc. We adopt $L = 35$~pc by scaling the distance to 6~kpc. The derived electron density is 13 cm$^{-3}$ as the highest and 7.3 cm$^{-3}$ on average, which is consistent with the result by Law et al. (2009). The total number of electrons in the plasma in the GCL is derived to be $N_{\rm e} = 7.7 \times 10^{61}$, and the total mass of the thermal plasma is $6.5 \times 10^{4} M_{\rm \solar}$. The thermal energy of the plasma is derived to be $E = (3/2)N_{\rm e}k_{\rm B}T_{\rm e} = 6.9 \times 10^{49}$ erg with the electron temperature $T_{\rm e} = 4360$ K. The magnetic field that gives the equivalent magnetic energy density $u_{\rm B}$ to the thermal energy density is estimated to be $B = 13 \mu$G for $n_{\rm e} = 7.3$ cm$^{-3}$ and $T_{\rm e} = 4360$ K, using the formula $B^{2}/8 \pi = (3/2) n_{\rm e} k_{\rm B} T_{\rm e}$. The magnetic field of the GCL estimated with Faraday rotation observations is $B = 5 \mu$G (Law et al. 2011). Therefore, the thermal component is the dominant energy density in the GCL over the magnetic-field one.

Thermal plasma is cooled by radiation. For the plasma with a temperature of $T_{\rm e} = 4360$ K, the cooling rate per unit volume per unit time is given by $\Lambda = 7 \times 10^{-25} n_{\rm e}^{2}$ erg cm$^{-3}$ s$^{-1}$ (Spitzer 1978). From the thermal energy density $u_{\rm t} = (3/2)n_{\rm e}k_{\rm B}T_{\rm e}$ and the cooling rate, the time scale of the energy dissipation is calculated to be $\tau_{\rm cool} = u_{\rm t} / \Lambda$. Adopting the density of 7.3 cm$^{-3}$, the dissipation time scale is $\tau_{\rm cool} = 5.2 \times 10^{3}$ yr. Assuming that the heating and cooling are balanced, the heating power is $P = E / \tau_{\rm cool} = 4.2 \times 10^{38}$ erg s$^{-1}$ or $1.1 \times 10^{5} L_{\rm \solar}$. Since the cooling time of the thermal plasma of the GCL is as short as $\tau_{\rm cool} = 5.2 \times 10^{3}$ yr, the plasma would be heated continuously at present by some heating source to sustain the current state of the GCL. The GCL is also detected by an emission line of CII 158 $\mu$m (Nakagawa et al. 1998). Since this line is a tracer of the photo-dissociation region enveloping HII regions and is created by ultraviolet photons, the thermal plasma of the GCL is ionized by ultraviolet radiation. All the derived physical parameters of the GCL, such as the size of roughly 100 pc, $n_{\rm e} = 7.3$ cm$^{-3}$, and $T_{\rm e} = 4360$ K, support that the GCL is a giant HII region.

Law et al. (2009) proposed the candidate sources to ionize the GCL as luminous star clusters, such as Quintuplet, Arches, and Nucleus Star Cluster. The luminosity of these clusters exceeds $1.1 \times 10^{5} L_{\rm \solar}$, which is the required luminosity to sustain the ionization of the GCL. However, as discussed in section 4.3, the GCL is likely to be located away from the center of the galaxy. Even if the GCL exists in the central region of the galaxy, the positions of Quintuplet and Arches are clearly outside the GCL (figure 7). Hence, the radiation source that ionize the GCL is unclear.

As discussed above, the GCL is a giant HII region filled with low-velocity thermal plasma. Accordingly, any models claiming that the GCL is a fast ($\sim 100$ km s$^{-1}$) jet ejected from the galactic center (e.g., Crocker et al. 2011) are ruled out. Sofue (1985) proposed a model that the poloidal magnetic field that penetrates the galactic plane is twisted by the galactic rotation, and that the GCL is a structure ejected from the galactic plane by the energy accumulated in the magnetic field. This models is also unlikely because this model requires high rotation velocity ($\sim 100$ km s$^{-1}$) and strong magnetic field ($\sim 100 \mu$G) but the observations showed slow the velocity of $\le 10$ km s$^{-1}$ and weak magnetic field of $5 \mu$G (Law et al. 2011) of the GCL.

Starburst galaxies such as NGC 3079 often have an ionized gas structure similar to the GCL (Cecil et al. 2001), which is considered to be formed by starburst radiation and outflows of hot gases. According to Veilleux et al. (2005), the mass of the ionized outflow observed for small-scale starburst is $10^{5} \sim 10^{6} M_{\rm \solar}$, and the energy is $10^{50} \sim 10^{54}$ erg. The estimated mass of the ionized gas of the GCL ($6.5 \times 10^{4} M_{\rm \solar}$) and the thermal energy ($6.9 \times 10^{49}$ erg) are close to the lower limit of the parameters of starburst galaxies. However, the gas velocity of the GCL is much slower than the typical value for normal starburst galaxies, 100 km s$^{-1}$, and models similar to the starburst or galactic center outflow (e.g., Veilleux et al. 2005) are probably not applicable to the GCL.

\section{Conclusion}
We observed the whole area of the GCL with the Yamaguchi 32 m radio telescope with 8.4 GHz continuum and H92$\alpha$ radio recombination line. The spatial intensity distribution of the radio recombination line shows two ridges in the eastern and western sides of the galactic center, forming a loop structure connected at a latitude of $1.2^{\circ}$. We find that the spatial distribution is in reasonable agreement with that of the 8.4 GHz continuum. This suggests that most of the continuum emission of the GCL is free-free emission from thermal plasma, and that the GCL is filled with the thermal plasma.

A detail comparison between the distributions of the radio continuum and recombination line suggests that neither the radio arc nor Sgr C would be physically related to the GCL. The radial velocity distribution of the radio recombination line is $-4$ to $+10$ km s$^{-1}$ across the GCL, which is much lower than the one expected from the galactic rotation. These properties suggest that the GCL is located in the foreground of the galactic center region. The distance between the GCL and the galactic center is estimated to be 2.6 kpc on the assumption that the velocity gradient of the GCL is caused by the galactic rotation. Since the GCL is filled with low-velocity thermal plasma, the models that propose high-velocity jets like starburst or rotation - acceleration by magnetic field as a formation mechanism of the GCL are not likely the case. The derived parameters of the GCL are as follows; the electron temperature of the thermal plasma in the GCL is $T_{\rm e} = 4360$ K, the electron density is $7.3$ cm$^{-3}$ on average, the mass of the thermal plasma is $6.5 \times 10^{4} M_{\rm \solar}$, the thermal energy of the plasma is $6.9 \times 10^{49}$ erg, the cooling time scale is $5.2 \times 10^{3}$ yr, and the required heating power is $1.1 \times 10^5 L_{\rm \solar}$. These parameters suggest that the GCL is a giant HII region.

\begin{ack}
We would like to thank Mr. Teppei Tamaki, Mr. Fumiaki Hayashi, Mr. Ryu Ueta, Ms. Yurie Suga, Ms. Miho Moriki, and the students in Yamaguchi University and Hokkaido University who supported the observation. We also would like to thank KDDI for their support in Yamaguchi 32 m radio telescope operation. One of the authors (KF) thanks Dr. Tomomi Shimoikura, Professor Masato Tsuboi, and Professor Hideki Uchiyama for useful discussion, and Professor Yoshiaki Sofue for his encouragement for this research, Dr. Masa Sakano, Wise Babel Ltd for proofreading the manuscript. We also thank the referee for useful comments.
\end{ack}


\end{document}